*Review*

# Crowdsourcing for Bioinformatics

Benjamin M. Good*, Andrew I. Su

Department of Molecular and Experimental Medicine, The Scripps Research Institute, La Jolla, CA, USA

**ABSTRACT**

**Motivation:** Bioinformatics is faced with a variety of problems that require human involvement. Tasks like genome annotation, image analysis, knowledge-base construction and protein structure determination all benefit from human input. In some cases people are needed in vast quantities while in others we need just a few with very rare abilities. Crowdsourcing encompasses an emerging collection of approaches for harnessing such distributed human intelligence. Recently, the bioinformatics community has begun to apply crowdsourcing in a variety of contexts, yet few resources are available that describe how these human-powered systems work and how to use them effectively in scientific domains.

**Results:** Here, we provide a framework for understanding and applying several different types of crowdsourcing. The framework considers two broad classes: systems for solving large-volume 'microtasks' and systems for solving high-difficulty 'megatasks'. Within these classes, we discuss system types including: volunteer labor, games with a purpose, microtask markets and open innovation contests. We illustrate each system type with successful examples in bioinformatics and conclude with a guide for matching problems to crowdsourcing solutions.

## 1 INTRODUCTION

Imagine having easy, inexpensive access to a willing team of millions of intelligent workers. What could you accomplish? Lakhani and colleagues produced thirty new sequence alignment algorithms that each improved upon the state of the art - in two weeks, for $6000 (Lakhani, et al., 2013). Others improved a 44-species multiple alignment (Kawrykow, et al., 2012), developed a new protein folding algorithm (Khatib, et al., 2011), produced accurate parasite counts for tens of thousands of images of infected blood cells (Luengo-Oroz, et al., 2012), and still others are attempting to translate the entire Web into every major language (http://duolingo.com). Crowdsourcing systems make these and many other monumental tasks possible. Here, we will explore what these systems are and how they are being applied in bioinformatics.

The term 'crowdsourcing' was coined in 2006 to describe "the act of taking a job traditionally performed by a designated agent (usually an employee) and outsourcing it to an undefined, generally large group of people in the form of an open call" (Howe, 2006). Now, it is used to describe an extremely wide range of ongoing activities. Here, we will focus on *systems for accomplishing directed work that requires human intelligence*. These kinds of crowdsourcing system are built to solve discrete tasks with clear endpoints. They are distinct from the related field of wikis in that they allow for top-down control over the work that is conducted and have a much greater diversity of potential work products. (For an extensive introduction to wikis in bioinformatics see the 2011 database issue of Nucleic Acids Research which includes ten wiki-related articles (Galperin and Fernandez-Suarez, 2012).)

Two other models that depend on the crowd, but not the crowd's intelligence, are distributed computing and online health research. Systems like Rosetta@home and the more-general purpose BOINC use the spare cycles of thousands of personal computers to advance research in bioinformatics, particularly protein-folding and docking simulations (Sansom, 2011). But, since they do not engage the minds of the crowd they will not be discussed here. In the medical domain, the term crowdsourcing is often used to describe large-scale patient data collection through online surveys. Personal genomics companies such as 23andme have surveyed their genotyped 'crowd' to enable many new discoveries in genetics (Do, et al., 2011; Tung, et al., 2011). In addition, a variety of initiatives have begun exploring the crowdsourcing of both patient-initiated and researcher-initiated (non-clinical) patient trials. Such "crowdsourced health research" is an important and growing area, but conceptually distinct from the crowdsourcing applications considered here. For a recent survey of the literature on this topic, see (Swan, 2012).

The tasks considered here are those that have historically been approached from an artificial intelligence perspective – where algorithms attempt to mimic human abilities (Sabou, et al., 2012). Now, crowdsourcing gives us access to a new methodology: "artificial artificial intelligence" (https://www.mturk.com/). The objective of this review is to highlight how, from a practical perspective based on recent successes, to use this new force to tackle difficult problems in biology.

We divide different types of crowdsourcing system into two major groups: those for solving 'microtasks' that are large in number but low in difficulty, and those for solving individually challenging 'megatasks'. Within these broad upper classes, crowdsourcing systems vary along many dimensions that can be used by system designers to tune them to meet specific goals. These dimensions include, but are not limited to: incentives, the nature of the work, and the approaches used to ensure high quality (Table 1).

---

*To whom correspondence should be addressed.

| Task Class | System Type | Conditions where appropriate | Examples | Explicit Incentive | Quality Control | Tools/Platforms |
|---|---|---|---|---|---|---|
| Micro | Volunteer | Tasks of interest to general public, very high task volume | Image classification (e.g. CellSlider, Galaxy Zoo) | None | R&A | Bossa, PyBossa |
| Micro | Casual Game | Access to game developers, very high task volume | Multiple sequence alignment, image classification (e.g. Phylo, MOLT) | Fun | R&A | None |
| Micro | Microtask Market | Access to sufficient funds for required volume of work | Image classification, text annotation (e.g. polyp classification for colon cancer detection) | Money | R&A | **Platforms**: Mechanical Turk, Clickworker, Microworkers, mobileworks **Meta services**: Crowdflower, Crowdsource **Tools:** Turkit, Crowdforge |
| Micro | Forced Labor | Control over a workflow that your target population needs to execute | Character recognition, linking drugs to clinical problems (e.g. ReCAPTCHA) | Completing another task of personal importance | R&A | None |
| Micro | Educational | Twin goals of education and task-completion | Genome annotation, document translation (e.g. DuoLingo) | Education | R&A | annotathon.org/ |
| Mega | Hard Game | Access to game developers, problem with solution quality function that can be tied to a game score | Protein folding, RNA structure design (e.g. Foldit, EteRNA) | Fun | Automatic scoring function, Peer review | None |
| Mega | Innovation Contest | Access to sufficient resources to provide desirable reward for solution. | Algorithm development (e.g. DTRA) | Money | Expert Review (possibly with automatic scoring) | Innocentive, TopCoder, Kaggle |

**Table 1. Crowdsourcing systems.** R&A = Redundancy and Aggregation. Types of crowdsourcing systems are displayed, from the top down, in roughly increasing order of difficulty per task and decreasing number of individual tasks that the system must solve. The easiest and most prolific are the character recognition microtasks of ReCAPTCHA while the most difficult are the innovation contests for megatasks like algorithm development.

Here, we will focus only on crowdsourcing approaches that are specifically relevant to common problems in bioinformatics. For broader reviews see "Crowdsourcing Systems on the World Wide Web" (Doan, et al., 2011), "Human-Computation: A Survey and Taxonomy of a Growing Field" (Quinn and Bederson, 2011) and "Crowd-Powered Systems" (Bernstein, 2012).

## 2 CROWDSOURCING MICROTASKS

Microtasks can be solved in a short amount of time (typically a few seconds) by any human that is capable of following a short series of instructions. In bioinformatics, microtasks often orient around image or text annotation. In these cases, crowdsourcing systems provide system designers with access to vast numbers of workers that, working in parallel, can collectively label enormous volumes of data in a short amount of time. These systems achieve high quality, typically as good or better than expert annotators, through extensive use of redundancy and aggregation. Annotation tasks are presented to multiple workers and their contributions are integrated, for example through voting, to arrive at the final solution.

The different crowdsourcing systems in this category vary primarily based on the kinds of incentives that are used to attract workers (Table 1). We now explore each in turn.

### 2.1 Volunteer (Citizen Science)

Perhaps the most surprisingly effective strategy for incentivizing large-scale labor in support of scientific objectives is simply to ask for volunteers. This pattern, often referred to as 'citizen science', dates back at least to the year 1900 when the annual Christmas bird counts were first organized by the National Audubon Society (Cohn, 2008). Now, it is exemplified by the Zooniverse project and its initial product Galaxy Zoo (Lintott, et al., 2008). Galaxy Zoo has successfully used the Web to tap into a willing community of contributors of previously unimaginable scale. Within the first ten days of its launch in July of 2007, the Galaxy Zoo website had captured 8 million morphological classifications of images of distant galaxies (Clery, 2011). After nine months, more than 100,000 people had contributed to the classification of more than 1,000,000 images - with an average of 38 volunteers viewing each image. Now, the Zooniverse project, in collaboration with Cancer Research UK, is moving into the biomedical domain with a project called CellSlider (http://www.cellslider.net).

In CellSlider, volunteers are presented with images of stained cell populations from cancer patient biopsies and asked to label the kinds and quantities of different cell types. In particular, volunteers seek out irregularly shaped cells that have been stained yellow based on the level of estrogen receptor (ER) expressed by the cell. Quantifying the amount of these 'cancer core' cells in a particular patient can help to ascertain the extent to which a treatment is helping the patient and can thus be used to help personalize and improve therapy. Launched on October 24, 2012 the initiative has not published its finding yet, but claimed to have analyzed 550,000 images in its first 3 months of operation.

## 2.2 Casual Games

Aside from simply relying on the altruistic urges of the audience, a growing number of crowdsourcing initiatives attempt to reward participation with fun. In these 'games with a purpose' (GWAP), microtasks are presented in the context of simple, typically web-based games (Ahn and Dabbish, 2008). (Note that we distinguish these from closely related games designed to solve very difficult problems presented below.) In these 'gamified' crowdsourcing systems, the participants earn points and advance through levels just like other games, but the objectives in each game are closely aligned with its higher-level purpose. In order to win, game players have to solve real-world problems with high quality and in large quantities.

Casual crowdsourcing games have been actively developed by the computer science community since the ESP Game emerged with great success for general-purpose image labeling in 2003 (Ahn and Dabbish, 2004). The first casual games for tasks in bioinformatics have only recently been published.

The first of these was Phylo, a game in which players help to improve large multiple sequence alignments by completing a series of puzzles representing dubious sections from pre-computed alignments (Kawrykow, et al., 2012). To complete a puzzle, players move Tetris-like, color-coded blocks representing nucleotides around until the computed alignment score reaches at least a predetermined level, with more points awarded for better alignments. These human-generated alignment sections are then integrated back into the full computationally generated alignments. In the first seven months of game-play, Phylo recruited more than 12,000 players who collectively completed more than 254,000 puzzles. When the alignment blocks from game-players were re-assembled, they resulted in improvements to more than 70% of the original alignments.

*2.2.1 Casual games for biological image annotation.* Following Phylo, independent research groups developed two games focused on the classification of images related to malaria infection concurrently. Mavandadi and colleagues describe a Web-based game called MOLT that challenges players to label red blood cells from thin blood smears as either infected or uninfected (Mavandadi, et al., 2012; Mavandadi, et al., 2012). Luengo-Oroz and colleagues present a game called MalariaSpot for counting malaria parasites in thick blood smears (Luengo-Oroz, et al., 2012). The similar approaches taken by both of these systems reflect consistent themes for microtask platforms; both systems aggregate the responses of multiple players (sometimes more than 20) to produce the annotation for each image and use images with known annotations to benchmark player performance. Using these techniques, both MOLT and MalariaSpot achieved expert-level performance on their respective tasks. Both systems share a vision of using their crowdsourcing approach to enable the rapid, accurate, and inexpensive annotation of medical images from regions without access to local pathologists in a process known as 'tele-pathology'. These systems are also envisioned to play a role in training both human pathologists and automated computer vision algorithms.

In addition to malaria diagnosis, a new image annotation game called EyeWire is currently collecting data about neurons in the retina (https://eyewire.org/). EyeWire players are helping to identify the structure and connectivity of individual neurons in three-dimensional images collected using serial block face scanning electron microscopy.

## 2.3 Microtask Markets

Microtask markets are probably the most well known and thoroughly used variety of crowdsourcing. Rather than attempting to use fun or altruism as incentives, these systems simply use cash rewards. Where a game like MalariaSpot provides points for each labeled image, a microtask market would allow contributors to earn a small amount of money for each unit of work. Within bioinformatics, microtask markets have so far been used for image and text annotation.

*2.3.1 Image annotation.* While microtask markets have enjoyed widespread use for general image annotation tasks since their inception, there are few published examples of applications in bioinformatics - though many are in progress. Nguyen and colleagues provide a prototypical example (Nguyen, et al., 2012). They describe the application of the Amazon Mechanical Turk (AMT) crowdsourcing service to detect polyps associated with colorectal cancer in images generated through computed tomographic colonography. Using the AMT, they paid crowd workers to label images of polyp candidates as either true or false. For each task (known as a 'HIT' for 'human intelligence task'), the workers were presented with 11 labeled training images to use to make their judgment on the test image. Workers were paid $0.01 for each image that they labeled. In the first of two replicate trials with nearly identical results, 150 workers collectively completed 5,360 tasks resulting in 20 independent assessments of each of 268 polyp candidates. This work was completed in 3.5 days at a total cost of $53.60 (plus some small overhead fees paid to Amazon). A straightforward voting strategy was used to combine the classifications made by multiple workers for each polyp candidate. The classifications generated by this system were then assessed based on agreement with expert classifications and compared with results from a machine learning algorithm. The results of the crowd-powered system and the machine learning system were not significantly different. Both systems produced an area under the receiver operating characteristic curve (AUC) close to 0.85. While this system did not improve upon the automated system, it demonstrated that minimally trained AMT workers could perform this expert-level task rapidly and with high quality. In subsequent work, the same research group reported significant improvements with a new system that integrated the automated predictions with those derived from crowdsourcing to produce an AUC of 0.91 on the same data (Wang, et al., 2011).

*2.3.2 Text annotation.* With much of world's biological and medical knowledge represented in text, Natural Language Processing (NLP) is a core component of research in bioinformatics. Many tasks in NLP require extensive amounts of expensive linguistic annotation. For example, NLP systems that detect concepts and relationships often need large corpuses of semantically tagged text for training (Kim, et al., 2003). In seeking a faster, less-expensive method for acquiring this data, the NLP community was among the first to explore the use of crowdsourcing for research purposes (Sabou, et al., 2012). Early work by Snow and colleagues demonstrated that expert-level text annotations could be collected "cheap and fast" using the AMT platform and also provided a pattern for correcting biases common to crowdsourcing systems (Snow, et al., 2008). While this and related work has achieved good results with common language tasks, biomedical text (with its more challenging vocabulary) is just beginning to be approached through crowdsourcing.

Yetisgen-Yildiz and colleagues demonstrated that AMT workers could produce effective annotations of medical conditions, medications, and laboratory tests within the text of clinical trial descriptions (Yetisgen-Yildiz, et al., 2010). Berger and colleagues also used the AMT to validate predicted gene-mutation relations in MEDLINE abstracts (Burger, et al., 2012). They found that the workers (paid $0.07/task) were easily recruited, responded quickly, and (as is typical of all crowdsourcing systems) displayed a wide range of abilities and response rates with the best-scoring worker producing an accuracy of 90.5% with respect to a gold standard on more than 1000 HITs. Using majority voting to aggregate the responses from each worker, they achieved an overall accuracy of 83.8% across all 1733 candidate gene-mutation relationships presented for verification. Finally, Zhai and colleagues recently showed that crowdsourcing could be used for detailed processing of the text from clinical trials announcements including: annotating named entities, validating annotations from other workers, and identifying linked attributes such as side effects of medications (Zhai, et al., 2012).

*2.3.3 Microtask platforms.* The AMT was the first and remains the leading microtask market, but there are a variety of other platforms emerging (Table 1). In addition, meta-services like Crowdflower help to address standard problems in microtask markets such as spammer identification, worker rating, and response aggregation. From the task-requestor perspective, the meta-services generally offer less control over the operation of the system but solve many common problems effectively. Aside from these services, a small but growing number of open source projects for working with crowdsourcing systems are now available. For example, see Turkit (Little, et al., 2010) and CrowdForge (Kittur, et al., 2011).

## 2.4 Forced Labor (Workflow Sequestration)

If altruism, fun, or money is not sufficient to motivate workers, it is sometimes possible to force them to work for free. This strategy has been used most effectively in the omnipresent ReCAPTCHA (Ahn, et al., 2008). ReCAPTCHA is a security system for websites that requires users to type in two words that they see in a distorted image. One word is known and thus used for verification that the user is a human (not a program) and the other is a scanned image of text that needs to be digitized. Since this task is very difficult to accomplish computationally, it provides a good way to defend against automated spammers. At the same time, it effectively forces hundreds of millions of Web users to work on large-scale optical character recognition tasks for free.

ReCAPTCHA uses the incentive to complete a task that is important to the user/worker (logging in to a website) to motivate them to complete a task that is important to the system designer (digitize books). McCoy and colleagues recently applied this pattern for clinical knowledge base construction (McCoy, et al., 2012). In this study, the 'crowd' consisted of the physicians in a large medical community, the incentive was to use an electronic health record system to prescribe medications, and the task was to capture links between medications and patient problems. To prescribe a medication, the physicians were required to link it to the associated clinical problem. Using this pattern, 867 clinicians created 239,469 problem-medication links in one year, including 41,203 distinct links. After filtering, the system yielded 11,166 distinct problem-medication pairs. Compared to expert review of the associated records, these links had a specificity of 99.6% and a sensitivity of 42.8%.

The success of this early study, conceptual articles that describe similar patterns (Hernández-Chan, et al., 2012) and the continued increase in adoption of electronic health record systems suggest that this approach will enjoy widespread application in the biomedical domain. Within bioinformatics, this kind of workflow sequestration is, so far, most commonly seen in educational settings.

## 2.5 Crowdsourcing and education

Genome annotation is a crucial activity in bioinformatics and is one that requires extensive human labor. With an ever-increasing supply of genomes to annotate, there is an effectively infinite amount of work to accomplish and, as this work is non-trivial, a need to train large numbers of students to accomplish it. Killing two birds with one stone, a number of annotation projects have incorporated the annotation of new sequences directly into the curriculum of undergraduate courses (Hingamp, et al., 2008). Using standard crowdsourcing mechanisms, redundancy and aggregation, as well as review by expert curators, these initiatives have generated thousands of high-quality annotations (Brister, et al., 2012).

From an economic perspective, this approach has the elegant property of simultaneously accomplishing the desired work and generating the capital needed to pay the workers. In this case the capital is the knowledge that they are acquiring by interacting with the system. The startup company DuoLingo employs this pattern on a massive scale by having millions of students learning foreign languages translate web documents (http://duolingo.com).

## 3 CROWDSOURCING MEGATASKS

In addition to rapidly completing large volumes of simple tasks, different incarnations of the crowdsourcing paradigm can be applied to solve individual tasks that might take weeks or even months of expert-level effort to complete. In these cases, the goal is to use crowdsourcing to seek out and enable the few talented individuals from a very large candidate population that might, through the heterogeneous skills and perspectives that they provide, be able to solve problems that continue to stymie traditional research organizations. This shift from high-volume tasks to high-difficulty tasks affords different requirements for successful crowdsourcing. Two approaches that have generated impressive successes in bioinformatics are hard games and innovation contests.

### 3.1 Hard Games

In contrast to casual games like MalariaSpot that are designed to complete large volumes of microtasks, the games discussed here provide players with access to small numbers of extremely challenging individual problems. While casual games tend toward what the gaming community describes as "grinding" where the players perform highly repetitive actions, hard games provide rich, interactive environments that promote long-term exploration and engagement with a challenge. Two such games have thus far been successful in bioinformatics, Foldit and EteRNA.

In Foldit, the goal of most games (or puzzles) is typically to find the three-dimensional conformation of a given protein structure that results in the lowest calculated free energy (Cooper, et al., 2010). To achieve this goal, players interact with a rich desktop game environment that builds upon the Rosetta structure prediction tool suite (Rohl, et al., 2004). In contrast to casual games in which players can play (and contribute solutions) within minutes, Foldit players must first advance through an extensive series of training levels that can take several hours to complete. These introductory levels systematically introduce increasingly complex game features that allow players to manipulate protein structures via both direct manipulation (dragging and twisting pieces of the protein) and through the execution of small optimization algorithms like 'wiggle'. Importantly, these training levels abstract the complex mechanics of protein folding into concepts that are accessible to lay game players.

Since its inception in 2008, Foldit has captured the attention of hundreds of thousands of players, some of whom have achieved remarkable scientific successes. Foldit players have outperformed some of the world's best automated structure prediction systems and aided in the solution of an important retroviral structure that had eluded specialists for decades (Khatib, et al., 2011). In addition to solving naturally occurring protein structures, players have recently succeeded in optimizing the design of engineered enzymes to achieve specific physico-chemical goals (Eiben, et al., 2012).

While these individual successes are impressive, the greater challenge remains to devise algorithms that fold proteins automatically. In addition to the visually-oriented puzzle interface, Foldit introduced a scripting system that allows players to compose automated workflows. These scripts string together multiple optimization widgets and may be used in combination with direct manipulation. In one of the most intriguing developments from this initiative, Foldit players used the provided scripting interface to collaboratively write folding algorithms that rival professionally designed solutions (Khatib, et al., 2011).

Following directly from Foldit's success, some of Foldit's creators have released a new game called EteRNA (http://eterna.cmu.edu). In EteRNA, the goal is to design an RNA molecule that will fold into a particular predefined shape. Design contests are run every week and the best designs are evaluated in the laboratory providing real-world feedback. This connection between the gamer community and the scientists behind the game has proven effective in recruiting tens of thousands of players - including a few star players that are not only producing valuable new designs but are also identifying new rules of RNA behavior (Koerner, 2012).

While much is made of the numbers of players to access these games, it is important to realize that only a very small fraction of these players contribute directly to any important advance. These games are portals for recruiting, engaging and enabling a small number of people with exceptional skills that would never normally have the opportunity to help solve these problems. In essence, these games are as much about discovering latent scientists as they are about making scientific discoveries (Good and Su, 2011).

Most of the players are not active scientists by trade and typically have little to no formal training. While most do not contribute directly to solutions, a few bring a different perspective that opens up an entirely new way of looking at and solving the problem. Such a diversity of human intelligence, if filtered and aggregated effectively, is a powerful and much sought after force.

## 3.2 Open Innovation Contests

Open innovation contests define particular challenges and invite anyone in the general public to submit candidate solutions. The solutions are evaluated and, if they meet the defined criteria, the best solutions are rewarded with cash prizes. The prizes, as well as the social prestige garnered by winning a large public contest, provide the key incentives driving participation in these initiatives. First pioneered by Innocentive, a 2001 spinoff of Eli Lilly meant to improve its research pipeline, a variety of platforms for operating these contests have recently emerged. Within bioinformatics, key open innovation platforms include Innocentive (which is used on a wide variety of tasks), TopCoder (for software development and algorithm design) and Kaggle (for data analysis).

As with games, contests make it possible to let enormous numbers of potential "solvers" each try out their unique abilities on the specified problem. In contrast to games, which require extensive, costly development time before any possible reward from the community might be attained; the up-front cost of running an innovation contest is comparatively small. If no one solves the posted problem, very little is lost by the problem-poster. Further, financial incentives are far easier to tune than game mechanics. The harder and more important the problem is, the bigger the offered bounty for its solution. Common prizes range from a few thousands dollars for small coding challenges that can be accomplished by individuals in their spare time to million dollar contests that can require large project teams and/or long-term time commitments.

Many successes in bioinformatics have already been attained at the lower end of the prize spectrum. As an example, Lakhani and colleagues recently assessed the potential of the TopCoder platform on a difficult sequence alignment problem (Lakhani, et al., 2013). To test the hypothesis that "big data biology is amenable to prize-based contests" they posted a challenge related to immune repertoire profiling on TopCoder with a prize pool of just $6,000. In the two weeks that the contest was run, 733 people participated and 122 submitted candidate solutions. In comparison to one prior "industry standard" (NCBI's MegaBlast), 30 of the submitted solutions produced more accurate alignments and all ran substantially faster. None of the participants in the competition was a professional computational biologist with most describing themselves as software developers. In addition to this academic study, industry representatives report extensive use of these small-scale coding competitions as part of the their bioinformatics R&D pipelines (Merriman, et al., 2012).

At the upper end of the prize spectrum, one of the first successful million-dollar contests lead to the discovery of a novel biomarker for Amyotrophic Lateral Sclerosis (Talan, 2011). Currently, groups such as Life Technologies and the U.S. Government's Defense Threat Reduction Agency (DTRA) are running million-dollar contests in the bioinformatics domain.

These examples highlight the potential of open innovation contests to focus the attention of large numbers of talented people on solving particular challenging problems. These systems offer solution seekers with an approach that can be highly cost-effective in recruiting such talent. As an example, Lakhani and colleagues estimate that contest participants spent approximately 2,684 hours working on their problem in total. Given a two-week time period and a total cost of $6,000, this is a remarkable amount of skilled labor and an incredibly short amount of time.

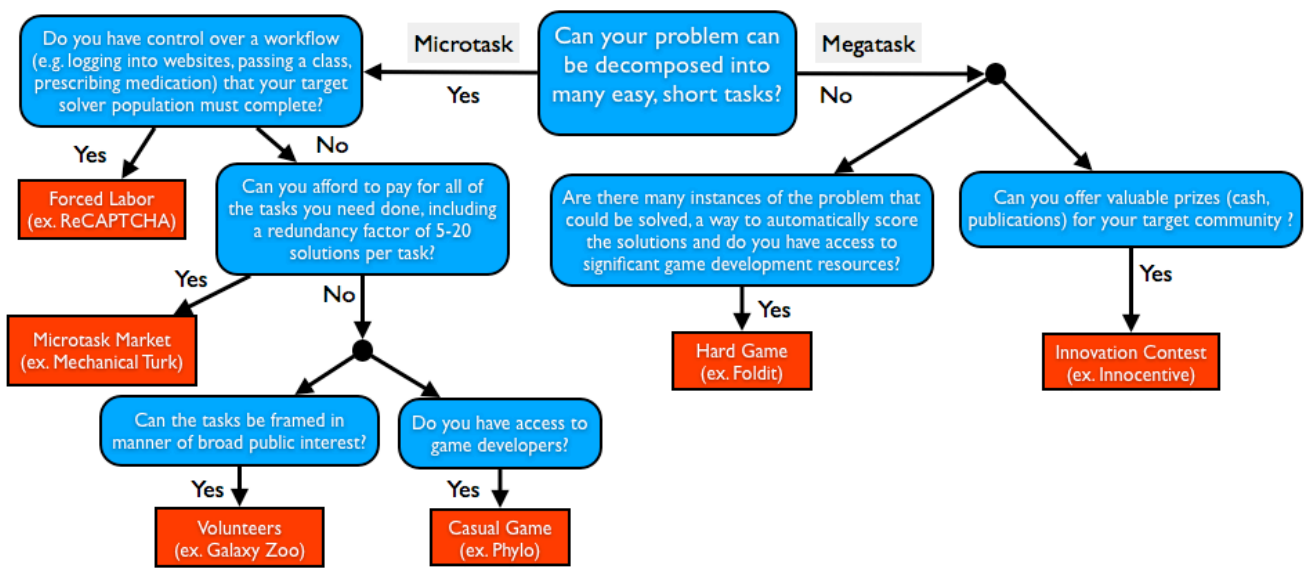

**Fig. 1. Crowdsourcing Decision Tree.** When considering a crowdsourcing approach, work through the tree from the top to identify approaches that may suit your particular challenge.

A variety of contests exist in the academic sphere, such as the long-running CASP for protein structure prediction and the recent series of challenges in systems biology operated by the DREAM initiative (Marbach, et al., 2012). For the most part, these contests remain distinct from other innovation contests in that they focus on recruiting submissions specifically from academics, using scientific publications as one form of incentive.

## 4 CHOOSING A CROWDSOURCING APPROACH

While diverse in their implementations and goals, the crowdsourcing systems described here each attempt to advance science by enabling the overwhelming majority of people that reside outside of the ivory tower to participate in the process (Cooper, 2013). How this process unfolds - how well it solves the problems at hand and how it influences the participants - depends deeply on the nature of each problem and the approach taken by system architects. While the diversity of potential tasks in bioinformatics renders a global rubric for composing crowdsourcing solutions unlikely, the examples presented in this review and organized in Table 1 suggest some general guidelines (Figure 1).

Highly granular, repetitive tasks such as image classification can be approached via volunteer initiatives, casual games, workflow sequestration and microtask markets. Games and direct volunteer labor are of most value when the number of required tasks is exceedingly large - too large to pay workers even small amounts per unit of work. The downsides of depending on volunteers or game players are that there is no guarantee that they will generate the required amount of labor and nearly all of the potentially substantial cost of building the crowdsourcing solution (the game, the website) must be paid up-front, before any possible benefit is attained. Depending on the task, workflow sequestration can be a very powerful approach as it not only effectively forces the required labor but can also be used to target very specific populations of workers. The downside is that the alignment of workflows with microtasks will likely not be possible in many cases. Finally, microtask markets have the benefit of offering system designers with an immediate workforce of massive scale as well as precise control of the nature and volume of their activities. The main negative aspect of microtask markets is that, because of the per-unit cost of the work, they do not have the capacity to scale up in the way that the other forms do.

When it comes to megatasks involving extended work and specialized skills, innovation contests and hard games can be considered. Among these, innovation contests are by far the most popular and generalizable framework. These systems have repeatedly produced solutions to very difficult problems in a variety of domains at comparatively tiny costs and we expect their use to continue to expand. Hard games, like Foldit, are fascinating for the potential scale, diversity and collaborative capacity of the gamer/solver population; however, these benefits are not guaranteed and come at a very high up-front cost in development time. Furthermore, it simply may not be possible to gamify many important tasks. The tasks most suited to approaches with hard games are those that have scoring functions, such as Foldit's free energy calculation, that can link performance in the game directly to the problem under study. Without such a mapping it will be very difficult to provide the players with the feedback they need to learn the problem space and thus become effective solvers.

Looking forward, the didactic division utilized here between systems for completing microtasks and those for solving megatasks will likely be blurred as new integrated systems arise that take advantage of key aspects of multiple forms of crowdsourcing (Bernstein, 2012). The emergent, community-driven processes that gave rise to Wikipedia offer some hints at what such future systems might look like (Kittur and Kraut, 2008). Such systems will have to promote the rapid formation of extended communities of participants that display a wide variety of skills and proclivities who come together to achieve a common high-level goal. For the moment, such problem solving communities remain difficult to generate and to sustain. But, as the science of crowdsourcing advances it will be increasingly possible for system architects to guide these collective intelligences into existence (Kittur, et al., 2011).

## 5   SOCIAL IMPACT

While we have focused primarily on the economic aspects of crowdsourcing, kinds of work and cost, there is another aspect that is important to consider. Crowdsourcing isn't just a new way of performing difficult computations rapidly and inexpensively; it represents a fundamental change in the way that scientific work is distributed within society. Recalling the original definition, crowdsourcing is a shift from work done in-house to work done in the open by anyone that is able. This means not only that we can often solve more problems more efficiently; it means that different people are solving them. As a result, there are both ethical concerns about worker exploitation that must be addressed and novel opportunities for societal side-benefits that are important to explore.

While some have expressed concern for the well-being of players of scientific crowdsourcing games (Graber and Graber, 2013), the majority of worry about exploitation is related to the workers in microtask markets. In some cases, people spend significant amounts of time earning wages that amount to less than $2/hour (Fort, et al., 2011). While problem-focused, resource-strapped researchers may rejoice at the opportunity to address the new scientific questions that this workforce makes possible, it is both socially responsible and vital for long term success to remain aware that there are people at the other end of the line completing these tasks. In fact many of the newer crowdsourcing companies, e.g. MobileWorks, now make worker conditions a top priority with guaranteed minimum wages and opportunity for advancement within their framework. Keeping worker satisfaction in mind should help encourage fair treatment but will also help designers come up with more effective crowdsourcing solutions. Paying workers well, building up long-term relationships with them and providing tasks that may provide them with benefits aside from any direct per-task reward in fun or money not only makes for a happier workforce, it makes for a far more powerful one (Kochhar, et al., 2010). While much is made of the power of our visual system in the context of crowdsourcing, our ability to learn is what separates us from the rest of the animal kingdom. Tapping into this innate ability and our strong desire to use it will produce crowdsourcing systems that not only solve scientific problems more effectively but, in the process, will end up producing many more scientists.

Before crowdsourcing models started to appear, only a very small fraction of society had any direct input into the advance of science. Consider protein folding. Foldit changed the number of people thinking about and working on protein folding problems from perhaps a few hundreds to hundreds of thousands. Consider also the new phenomenon of 'crowdfunding' (Wheat, et al., 2013). Now members of the public, not just members of government grant review panels have a vote in what science is funded.

The majority of Foldit players won't contribute to an important advance, but some will. Perhaps more importantly, Foldit players and contributors to the various other crowdsourcing initiatives discussed here are much more cognizant of these scientific problems than they ever were before. If fostered effectively by system architects, a new crowdsourcing-generated awareness will improve how the general public perceives science, how they vote and how they encourage future generations.

Taken together, the different manifestations of the crowdsourcing paradigm open up many new avenues for scientific exploration. From the high-throughput annotation of millions of images to the one-off introduction of a novel twist on RNA structure design by a librarian, these new systems are expanding scientific problem-solving capacity in unpredictable ways. To take advantage of these new ways of accomplishing work takes both openness and, in some cases, some amount of humility. You must be willing to share your greatest problems and step aside to let others help you solve them.

### ACKNOWLEDGEMENTS

Thanks to Hassan Masum, Sébastien Boivert, Jacques Corbeil, Mark Wilkinson, Attila Csordas and Twitter correspondents for helpful comments on an early draft of this manuscript.

*Funding*: NIH grants GM083924 and GM089820 to A.S.

### REFERENCES

Ahn, L.v. and Dabbish, L. (2004) Labeling images with a computer game. ACM Press, pp. 319-326.

Ahn, L.v. and Dabbish, L. (2008) Designing games with a purpose, *Commun. ACM*, **51**, 58-67.

Ahn, L.V*., et al.* (2008) reCAPTCHA: Human-Based Character Recognition via Web Security Measures, *Science*, **321**, 1465-1468.

Bernstein, M.S. (2012) Crowd-powered systems. *Electrical Engineering and Computer Science*. Massachusetts Institute of Technology.

Brister, J.R., Le Mercier, P. and Hu, J.C. (2012) Microbial virus genome annotation- Mustering the troops to fight the sequence onslaught., *Virology*, **434**, 175-180.

Burger, J*., et al.* (2012) Validating Candidate Gene-Mutation Relations in MEDLINE Abstracts via Crowdsourcing. In Bodenreider, O. and Rance, B. (eds), *Data Integration in the Life Sciences*. Springer Berlin Heidelberg, pp. 83-91.

Clery, D. (2011) Galaxy evolution. Galaxy zoo volunteers share pain and glory of research, *Science*, **333**, 173-175.

Cohn, J.P. (2008) Citizen Science: Can Volunteers Do Real Research?, *BioScience*, **58**, 192.

Cooper, C. (2013) The most stressful science problem. *Scientific American Blog*.

Cooper, S*., et al.* (2010) Predicting protein structures with a multiplayer online game, *Nature*, **466**, 756-760.

Do, C.B*., et al.* (2011) Web-based genome-wide association study identifies two novel loci and a substantial genetic component for Parkinson's disease., *PLoS genetics*, **7**, e1002141.

Doan, A., Ramakrishnan, R. and Halevy, A.Y. (2011) Crowdsourcing systems on the World-Wide Web, *Communications of the ACM*, **54**, 86.

Eiben, C.B*., et al.* (2012) Increased Diels-Alderase activity through backbone remodeling guided by Foldit players., *Nature biotechnology*, **30**, 190-192.

Fort, K.n., Adda, G. and Cohen, K.B. (2011) Amazon Mechanical Turk: Gold Mine or Coal Mine?, *Computational Linguistics*, **37**, 413-420.

Galperin, M.Y. and Fernandez-Suarez, X.M. (2012) The 2012 Nucleic Acids Research Database Issue and the online Molecular Biology Database Collection, *Nucleic Acids Research*, **40**, D1-8.

Good, B. and Su, A. (2011) Games with a scientific purpose, *Genome Biology*, **12**, 135.

Graber, M.A. and Graber, A. (2013) Internet-based crowdsourcing and research ethics: the case for IRB review, *J Med Ethics*, **39**, 115-118.

Hernández-Chan, G*., et al.* (2012) Knowledge Acquisition for Medical Diagnosis Using Collective Intelligence., *Journal of medical systems*, **36**, 5-9.

Hingamp, P*., et al.* (2008) Metagenome annotation using a distributed grid of undergraduate students., *PLoS biology*, **6**, e296.

Howe, J. (2006) The Rise of Crowdsourcing. *Wired*.

Kawrykow, A*., et al.* (2012) Phylo: a citizen science approach for improving multiple sequence alignment, *PloS one*, **7**, e31362.

Khatib, F*., et al.* (2011) Algorithm discovery by protein folding game players, *Proceedings of the National Academy of Sciences of the United States of America*, **108**, 18949-18953.

Khatib, F*., et al.* (2011) Crystal structure of a monomeric retroviral protease solved by protein folding game players, *Nat Struct Mol Biol*, **18**, 1175-1177.

Kim, J.-D*., et al.* (2003) GENIA corpus, A semantically annotated corpus for bio-textmining, *Bioinformatics*, **19**, i180-i182.

Kittur, A. and Kraut, R.E. (2008) Harnessing the wisdom of crowds in wikipedia: quality through coordination. *Proceedings of the 2008 ACM conference on Computer supported cooperative work*. ACM, San Diego, CA, USA, pp. 37-46.

Kittur, A*., et al.* (2011) CrowdForge: crowdsourcing complex work. *Proceedings of the 24th annual ACM symposium on User interface software and technology*. ACM, Santa Barbara, California, USA, pp. 43-52.

Kochhar, S., Mazzocchi, S. and Paritosh, P. (2010) The anatomy of a large-scale human computation engine. *Proceedings of the ACM SIGKDD Workshop on Human Computation*. ACM, Washington DC, pp. 10-17.

Koerner, B.I. (2012) New Videogame Lets Amateur Researchers Mess With RNA. *Wired Science*.

Lakhani, K.R*., et al.* (2013) Prize-based contests can provide solutions to computational biology problems, *Nat Biotech*, **31**, 108-111.

Lintott, C.J*., et al.* (2008) Galaxy Zoo: morphologies derived from visual inspection of galaxies from the Sloan Digital Sky Survey★, *Monthly Notices of the Royal Astronomical Society*, **389**, 1179-1189.

Little, G*., et al.* (2010) TurKit. *Proceedings of the 23nd annual ACM symposium on User interface software and technology - UIST '10*. ACM Press, New York, New York, USA, pp. 57.

Luengo-Oroz, M.A., Arranz, A. and Frean, J. (2012) Crowdsourcing Malaria Parasite Quantification: An Online Game for Analyzing Images of Infected Thick Blood Smears, *Journal of Medical Internet Research*, **14**, e167.

Marbach, D*., et al.* (2012) Wisdom of crowds for robust gene network inference., *Nature methods*, **9**, 796-804.

Mavandadi, S*., et al.* (2012) Distributed medical image analysis and diagnosis through crowd-sourced games: a malaria case study, *PloS one*, **7**, e37245.

Mavandadi, S*., et al.* (2012) Crowd-sourced BioGames: managing the big data problem for next-generation lab-on-a-chip platforms, *Lab Chip*, **12**, 4102-4106.

McCoy, A.B*., et al.* (2012) Development and evaluation of a crowdsourcing methodology for knowledge base construction: identifying relationships between clinical problems and medications., *Journal of the American Medical Informatics Association : JAMIA*, **19**, 713-718.

Merriman, B., R D Team, I.T. and Rothberg, J.M. (2012) Progress in Ion Torrent semiconductor chip based sequencing., *Electrophoresis*, **33**, 3397-3417.

Nguyen, T.B*., et al.* (2012) Distributed human intelligence for colonic polyp classification in computer-aided detection for CT colonography., *Radiology*, **262**, 824-833.

Quinn, A.J. and Bederson, B.B. (2011) Human-Computation: A survey and Taxonomy of a Growing Field. *CHI '11 SIGCHI Conference on Human Factors in Computing Systems*. ACM Press, New York, New York, USA, pp. 1403-1412.

Rohl, C.A*., et al.* (2004) Protein structure prediction using Rosetta, *Methods Enzymol*, **383**, 66-93.

Sabou, M., Bontcheva, K. and Scharl, A. (2012) Crowdsourcing research opportunities. *Proceedings of the 12th International Conference on Knowledge Management and Knowledge Technologies - i-KNOW '12*. ACM Press, New York, New York, USA, pp. 1.

Sansom, C. (2011) The power of many, *Nature Biotechnology*, **29**, 201-203.

Snow, R*., et al.* (2008) Cheap and fast---but is it good?: evaluating non-expert annotations for natural language tasks, 254-263.

Swan, M. (2012) Crowdsourced health research studies: an important emerging complement to clinical trials in the public health research ecosystem., *Journal of medical Internet research*, **14**, e46.

Talan, J. (2011) A Million Dollar Idea, Potential Biomarker for ALS, *Neurology Today*, **11**, 1.

Tung, J.Y*., et al.* (2011) Efficient replication of over 180 genetic associations with self-reported medical data., *PloS one*, **6**, e23473.

Wang, S*., et al.* (2011) Fusion of machine intelligence and human intelligence for colonic polyp detection in CT colonography. *2011 IEEE International Symposium on Biomedical Imaging: From Nano to Macro*. IEEE, pp. 160-164.

Wheat, R.E*., et al.* (2013) Raising money for scientific research through crowdfunding, *Trends in Ecology & Evolution*, **28**, 71-72.

Yetisgen-Yildiz, M*., et al.* (2010) Preliminary experience with Amazon's Mechanical Turk for annotating medical named entities. *CSLDAMT '10 Proceedings of the NAACL HLT 2010 Workshop on Creating Speech and Language Data with Amazon's Mechanical Turk*. Association for Computational Linguistics, Stroudsburg, PA, USA, pp. 180-183.

Zhai, H*., et al.* (2012) Cheap, Fast, and Good Enough for the Non-biomedical Domain but is It Usable for Clinical Natural Language Processing? Evaluating Crowdsourcing for Clinical Trial Announcement Named Entity Annotations. *2012 IEEE Second International Conference on Healthcare Informatics, Imaging and Systems Biology*. IEEE, pp. 106-106.